\documentclass[english,conference, twocolumn]{IEEEtran}
\usepackage[T1]{fontenc}
\usepackage[latin9]{inputenc}
\usepackage{verbatim}
\usepackage{amsthm}
\usepackage{amsmath}
\usepackage{amssymb}
\usepackage{esint}

\makeatletter
\theoremstyle{definition}
\newtheorem{assumption}{Assumption}
  \theoremstyle{plain}
  \newtheorem{thm}{\protect\theoremname}
  \theoremstyle{remark}
  \newtheorem{rem}{\protect\remarkname}

\usepackage{cite}
\usepackage{accents}
\IEEEoverridecommandlockouts

\makeatother

\usepackage{babel}
\providecommand{\remarkname}{Remark}
\providecommand{\theoremname}{Theorem}

\begin{document}

\title{Concurrent learning-based approximate optimal regulation%
\thanks{Rushikesh Kamalapurkar, Patrick Walters, and Warren Dixon are with
the Department of Mechanical and Aerospace Engineering, University
of Florida, Gainesville, FL, USA. Email: \{rkamalapurkar, walters8,
wdixon\}@ufl{}.edu.%
}%
\thanks{This research is supported in part by NSF award numbers 0547448, 0901491
and 1161260 and ONR grant number N00014-13-1-0151. Any opinions, findings
and conclusions or recommendations expressed in this material are
those of the author(s) and do not necessarily reflect the views of
the sponsoring agency. %
}}

\author{Rushikesh Kamalapurkar, Patrick Walters, and Warren Dixon }
\maketitle
\begin{abstract}
In deterministic systems, reinforcement learning-based online approximate
optimal control methods typically require a restrictive persistence
of excitation (PE) condition for convergence. This paper presents
a concurrent learning-based solution to the online approximate optimal
regulation problem that eliminates the need for PE. The development
is based on the observation that given a model of the system, the
Bellman error, which quantifies the deviation of the system Hamiltonian
from the optimal Hamiltonian, can be evaluated at any point in the
state space. Further, a concurrent learning-based parameter identifier
is developed to compensate for parametric uncertainty in the plant
dynamics. Uniformly ultimately bounded (UUB) convergence of the system
states to the origin, and UUB convergence of the developed policy
to the optimal policy are established using a Lyapunov-based analysis.
\end{abstract}

\section{Introduction}

Reinforcement learning (RL) enables a cognitive agent to learn desirable
behavior from interactions with its environment. In control theory,
the desirable behavior is typically quantified using a cost function,
and the control problem is formulated as the desire to find the optimal
policy that minimizes the cumulative cost. %
Recently, various RL-based techniques have been developed to approximately
solve optimal control problems for continuous-time and discrete-time
deterministic systems \cite{Bhasin.Kamalapurkar.ea2013a,Vamvoudakis2009,Vamvoudakis2010,Vrabie2007,Vrabie2009,Dierks2009,Dierks2009a,Zhang.Cui.ea2011,Doya2000,Al-Tamimi2007a,Al-Tamimi2008,Padhi2006,Mehta.Meyn2009}.
The approximate solution is facilitated via value function approximation,
where the value function is approximated using a linear-in-the-parameters
(LP) approximation, and the optimal policy is computed based on the
estimated value function.

Methods that seek an online solution to the optimal control problem,
(cf., \cite{Vrabie2007,Bhasin.Kamalapurkar.ea2013a}) are structurally
similar to adaptive control schemes. In adaptive control, the estimates
for the uncertain parameters in the plant model are updated using
the current tracking error as the performance metric, whereas, in
online RL-based techniques, estimates for the uncertain parameters
in the value function are updated using the Bellman error (BE) as
the performance metric. Convergence of online RL-based techniques
to the optimal solution is analogous to parameter convergence in adaptive
control.

Parameter convergence has been a focus of research in adaptive control
for several decades. It is common knowledge that the least squares
and gradient descent-based update laws generally require persistence
of excitation (PE) in the system state for convergence of the parameter
estimates. Modification schemes such as projection algorithms, $\sigma-$modification,
and $e-$modification are used to guarantee boundedness of parameter
estimates and overall system stability. However, these modification
schemes do not guarantee parameter convergence unless the PE condition,
which is often impossible to verify online, is satisfied \cite{Narendra1987,Narendra1989,Ioannou1996,Sastry1989a}.

As recently shown in results such as \cite{Chowdhary.Johnson2011a}
and \cite{Chowdhary.Yucelen.ea2012}, concurrent learning methods
can be used to guarantee parameter convergence in adaptive control
without relying on the PE condition. Concurrent learning relies on
recorded state information along with current state measurements to
update the parameter estimates. Learning from recorded data is effective
since it is based on the model error, which is closely related to
the parameter estimation error. The key concept that enables the computation
of the model error from past recorded data is that the model error
can be computed if the state derivative is known, and the state derivative
can be accurately computed at a past recorded data point using numerical
smoothing techniques \cite{Chowdhary.Johnson2011a,Chowdhary.Yucelen.ea2012}.

In RL-based approximate online optimal control, parameter estimates
are updated based on the BE along the state trajectories. Such weight
update strategies create two challenges for analyzing convergence.
The system states need to be PE for parameter convergence, and the
policy, which is based on the estimated weights, needs to regulate
the system states to a neighborhood around the desired goal so the
information around the desired trajectory can be used to learn the
value function. For example, in an infinite horizon regulation problem,
if the policy does not regulate the system states to a neighborhood
around the origin, the optimal value function (and hence, the optimal
policy) near the origin can not be learned, defeating one of the control
objectives. These challenges are typically addressed by adding an
exploration signal to the control input (cf. \cite{Mehta.Meyn2009,Vrabie2007,Sutton1998})
to ensure sufficient exploration in the desired region of the state
space. However, no analytical methods exist to compute the appropriate
exploration signal for nonlinear systems.

In this paper, the aforementioned challenges are addressed for an
infinite horizon optimal regulation problem on a nonlinear, control
affine plant with LP uncertainties in the drift dynamics by observing
that if the system dynamics are known, the BE can be computed at any
desired point in the state space. Unknown parameters in the value
function can therefore be adjusted based on least square minimization
of the BE evaluated at any number of desired points in the state space.
For example, in an infinite horizon regulation problem, the BE can
be computed at sampled points uniformly distributed in a neighborhood
around the origin of the state space. The results of this paper indicate
that convergence of the unknown parameters in the value function is
guaranteed provided the selected points satisfy a rank condition that
is weaker than the PE condition. Since the BE can be evaluated at
any desired point in the state space, sufficient exploration can be
achieved by appropriately selecting the points in a desired neighborhood.

If the system dynamics are partially unknown, an approximation to
the BE can be evaluated at any desired point in the state space based
on an estimate of the system dynamics. In this paper, a concurrent
learning-based parameter estimator is developed to exponentially identify
the unknown parameters in the system model, and the parameter estimates
are used to compute an approximation to the BE. The unknown parameters
in the value function are updated based on the approximate BE, and
uniformly ultimately bounded (UUB) convergence of the system states
to the origin and UUB convergence of the parameter estimates (and
hence, UUB convergence of the developed policy to the optimal policy)
is established using a Lyapunov-based analysis.

\section{Problem Formulation}

Consider a control affine nonlinear dynamic system 
\begin{equation}
\dot{x}\left(t\right)=f\left(x\left(t\right)\right)+g\left(x\left(t\right)\right)\hat{u}\left(t\right),\: t\in\left(0,\infty\right],\label{eq:dynamics}
\end{equation}
where $x\in\mathbb{R}^{n}$ denotes the system state, $\hat{u}\in\mathbb{R}^{m}$
denotes the control input, $f:\mathbb{R}^{n}\to\mathbb{R}^{n}$ denotes
the drift dynamics, and $g:\mathbb{R}^{n}\to\mathbb{R}^{n\times m}$
denotes the control effectiveness matrix. The objective is to solve
the infinite horizon optimal regulation problem online, i.e., to find
the optimal policy $u^{*}:\mathbb{R}^{n}\to\mathbb{R}^{m}$ defined
as 
\begin{equation}
u^{*}\triangleq\underset{\hat{u}\in U}{\underset{\hat{u}:\mathbb{R}^{n}\to\mathbb{R}^{m}}{arg\: min}}\intop_{t_{0}}^{\infty}r\left(x\left(\tau\right),\hat{u}\left(x\left(\tau\right)\right)\right)d\tau,\label{eq:u*def}
\end{equation}
while regulating the system states to the origin. In (\ref{eq:u*def}),
$U$ denotes the set of admissible state feedback policies, and $r:\mathbb{R}^{n}\times\mathbb{R}^{m}\to\left[0,\infty\right)$
denotes the instantaneous cost defined as 
\[
r\left(x,\hat{u}\right)\triangleq x^{T}Qx+\hat{u}^{T}R\hat{u},
\]
where $Q\in\mathbb{R}^{n\times n}$ and $R\in\mathbb{R}^{m\times m}$
are constant positive definite symmetric matrices. The class of nonlinear
systems considered in this paper is characterized by the following
assumption.
\begin{assumption}
\label{ass:fg}The drift dynamics $f$ is an unknown, LP locally Lipschitz
function such that $f\left(0\right)=0$, and the control effectiveness
matrix $g$ is a known, bounded locally Lipschitz function.
\end{assumption}
A closed-form solution to the optimal control problem is formulated
in terms of the optimal value function $V^{*}:\mathbb{R}^{n}\to\left[0,\infty\right)$
defined as
\begin{equation}
V^{*}\left(x_{0}\right)\triangleq\underset{\hat{u}\in U}{\underset{\hat{u}:\mathbb{R}^{n}\to\mathbb{R}^{m}}{min}}\intop_{t_{0}}^{\infty}r\left(x\left(\tau\right),\hat{u}\left(x\left(\tau\right)\right)\right)d\tau,\:\forall x_{0}\in\mathbb{R}^{n},\label{eq:V*def}
\end{equation}
where $x\left(\tau\right),\:\tau\in[t_{0},\infty)$ denote the trajectory
of (\ref{eq:dynamics}) with the feedback control law $\hat{u}\left(x\left(\tau\right)\right)$
and the initial condition $x\left(t_{0}\right)=x_{0}$. Assuming that
$V^{*}$ is continuously differentiable, and $V^{*}\left(0\right)=0,$
the optimal control law can be determined as 
\[
u^{*}=-\frac{1}{2}R^{-1}g^{T}\left(\nabla_{x}V^{*}\right)^{T},
\]
where $\nabla_{x}$ denotes the partial derivative with respect to
$x$. 

The optimal value function can be obtained by solving the corresponding
Hamilton-Jacobi-Bellman (HJB) equation
\begin{equation}
\nabla_{x}V^{*}\left(f+gu^{*}\right)+x^{T}Qx+u^{*T}Ru^{*}=0.\label{eq:HJB}
\end{equation}
Analytical solution of the HJB equation is generally infeasible; hence,
an approximate solution is sought. An approximate solution based on
minimizing the BE is facilitated by replacing $V^{*}$ and $u^{*}$
in (\ref{eq:HJB}) by their respective subsequently defined estimates
$\hat{V}$ and $\hat{u}$ to compute the BE $\delta\left(\hat{V},x,\hat{u}\right)\in\mathbb{R}$
as 
\begin{equation}
\delta=\nabla_{x}\hat{V}\left(f+g\hat{u}\right)+x^{T}Qx+\hat{u}^{T}R\hat{u}.\label{eq:delta1}
\end{equation}
The control objective is achieved by simultaneously adjusting the
estimates $\hat{V}$ and $\hat{u}$ to minimize the BE evaluated along
the trajectory $x\left(t\right)$. The BE depends on the drift dynamics
$f$. Since the drift dynamics are unknown, an adaptive system identifier
is developed in the following section.

\section{\label{sec:System-Identifier}System Identification}

Let $f\left(x\right)=Y\left(x\right)\theta^{*}$ be the linear parametrization
of the function $f$, where $Y:\mathbb{R}^{n}\to\mathbb{R}^{n\times p}$
is the regression matrix and $\theta^{*}\in\mathbb{R}^{p}$ is the
vector of constant unknown parameters. Let $\hat{f}:\mathbb{R}^{n}\times\mathbb{R}^{p}\to\mathbb{R}^{n}$
be an estimate of the unknown function $f$ defined as $\hat{f}\left(x,\hat{\theta}\right)\triangleq Y\left(x\right)\hat{\theta}$,
where $\hat{\theta}\left(t\right)\in\mathbb{R}^{n}$ is the vector
of parameter estimates, To estimate the drift dynamics, an identifier
is designed as
\begin{equation}
\dot{\hat{x}}=\hat{f}+g\hat{u}+k_{x}\tilde{x},\label{eq:observer}
\end{equation}
where the state estimation error $\tilde{x}$ is defined as $\tilde{x}\triangleq x-\hat{x}$
and $k_{x}\in\mathbb{R}^{n\times n}$ is a positive definite, constant
diagonal observer gain matrix. From (\ref{eq:dynamics}) and (\ref{eq:observer})
the identification error dynamics can be derived as
\begin{equation}
\dot{\tilde{x}}=Y\tilde{\theta}-k_{x}\tilde{x},\label{eq:xtildedot}
\end{equation}
where $\tilde{\theta}$ is the parameter identification error defined
as $\tilde{\theta}\triangleq\theta^{*}-\hat{\theta}.$

\subsection{Concurrent learning-based parameter update}

In traditional adaptive control, convergence of the estimates $\hat{\theta}\left(t\right)$
to their true values $\theta^{*}$ is ensured by using a PE condition\cite{Ioannou1996,Sastry1989a,Narendra1989}.
To ensure convergence without the PE condition, a concurrent learning-based
approach can be employed \cite{Chowdhary.Yucelen.ea2012,Chowdhary.Johnson2011a}.
The following observability assumption relaxes the PE condition that
is required for parameter convergence in adaptive control.
\begin{assumption}
\cite{Chowdhary.Yucelen.ea2012,Chowdhary.Johnson2011a}\label{ass:conccond}
There exists a finite set of time instances $\left\{ t_{j}\mid j=1,\cdots,M\right\} $
such that 
\begin{equation}
rank\left(\sum_{j=1}^{M}Y_{j}^{T}Y_{j}\right)=p,\label{eq:RankCond}
\end{equation}
where $Y_{j}=Y\left(x\left(t_{j}\right)\right).$
\end{assumption}
The condition in (\ref{eq:RankCond}) is satisfied as long as the
system states are exciting over a finite period of time, and hence,
is weaker than the PE condition. Furthermore, unlike the PE condition,
the rank condition in (\ref{eq:RankCond}) can be verified online
since it is a function of past states. To design the concurrent learning-based
parameter update law, time instances $\left\{ t_{j}\mid j=1,\cdots,M\right\} $
are selected such that the condition in (\ref{eq:RankCond}) holds,
and the states $\left\{ x_{j}\triangleq x\left(t_{j}\right)\mid j=1,\cdots,M\right\} $
and the corresponding control values $\left\{ \hat{u}_{j}\triangleq\hat{u}\left(t_{j}\right)\mid j=1,\cdots,M\right\} $
are recorded in a history stack. The update law is then designed as
\begin{equation}
\dot{\hat{\theta}}=\Gamma_{\theta}Y^{T}\tilde{x}+\Gamma_{\theta}k_{\theta}\sum_{j=1}^{M}Y_{j}^{T}\left(\dot{x}_{j}-g_{j}\hat{u}_{j}-Y_{j}\hat{\theta}\right),\label{eq:ThetaUpdate}
\end{equation}
where $g_{j}=g\left(x_{j}\right)$, $\Gamma_{\theta}\in\mathbb{R}^{p\times p}$
is a constant positive definite adaptation gain matrix, and $k_{\theta}$
is a constant positive concurrent learning gain. The update law in
(\ref{eq:ThetaUpdate}) depends on the unknown state derivative $\dot{x}_{j}=\dot{x}\left(t_{j}\right).$
However, since the state derivative is from recorded data, numerical
smoothing techniques based on past and future data can be used to
obtain good estimates of the derivative. In the presence of derivative
estimation errors, the parameter estimation errors can be shown to
be UUB, where the size of the ultimate bound depends on the error
in the derivative estimate\cite{Chowdhary.Yucelen.ea2012}. From (\ref{eq:dynamics})
and the definitions of $\tilde{\theta}$ and $\hat{f}$, the bracketed
term in (\ref{eq:ThetaUpdate}), can be expressed as $\dot{x}_{j}-g_{j}\hat{u}_{j}-Y_{j}\hat{\theta}=Y_{j}\tilde{\theta}$
and the parameter update law in (\ref{eq:ThetaUpdate}) can be expressed
as 
\begin{equation}
\dot{\hat{\theta}}=\Gamma_{\theta}Y^{T}\tilde{x}+\Gamma_{\theta}k_{\theta}\left(\sum_{j=1}^{M}Y_{j}^{T}Y_{j}\right)\tilde{\theta}.\label{eq:ThetaHatDotAlt}
\end{equation}

\subsection{Convergence analysis}

Let $V_{0}:\mathbb{R}^{n+p}\to\left[0,\infty\right)$ be a positive
definite continuously differentiable candidate Lyapunov function defined
as 
\begin{align}
V_{0} & \triangleq\frac{1}{2}\tilde{x}^{T}\tilde{x}+\frac{1}{2}\tilde{\theta}^{T}\Gamma_{\theta}^{-1}\tilde{\theta}.\label{eq:V0}
\end{align}
The following bounds on the Lyapunov function can be established:
\begin{equation}
\underline{v}\left\Vert z\right\Vert \leq V_{0}\leq\overline{v}\left\Vert z\right\Vert ,\label{eq:V0bound}
\end{equation}
where $\underline{v}=\frac{1}{2}min\left(1,\underline{\gamma}\right)$
and $\overline{v}=\frac{1}{2}max\left(1,\overline{\gamma}\right)$
are positive known constants, $z\triangleq\left[\tilde{x}^{T},\:\:\tilde{\theta}^{T}\right]^{T}\in\mathbb{R}^{n+p},$
and $\underline{\gamma},\overline{\gamma}\in\mathbb{R}$ denote the
minimum and the maximum eigenvalues of the matrix $\Gamma_{\theta}^{-1}$.

The time derivative of the Lyapunov function is given by
\begin{align}
\dot{V}_{0} & =\tilde{x}^{T}\dot{\tilde{x}}-\tilde{\theta}^{T}\Gamma_{\theta}^{-1}\dot{\hat{\theta}}.\label{eq:V0dot1}
\end{align}
Using (\ref{eq:xtildedot}) and (\ref{eq:ThetaHatDotAlt}), the Lyapunov
derivative in (\ref{eq:V0dot1}) can be expressed as 
\begin{equation}
\dot{V}_{0}=-\tilde{x}^{T}k_{x}\tilde{x}-\tilde{\theta}^{T}k_{\theta}\left(\sum_{j=1}^{M}Y_{j}^{T}Y_{j}\right)\tilde{\theta}.\label{eq:V0dot2}
\end{equation}
Let $\underline{y}\in\mathbb{R}$ be the minimum eigenvalue of $\left(\sum_{j=1}^{M}Y_{j}^{T}Y_{j}\right)$.
Since $\left(\sum_{j=1}^{M}Y_{j}^{T}Y_{j}\right)$ is symmetric and
positive semi-definite, (\ref{eq:RankCond}) can be used to conclude
that it is also positive definite, and hence $\underline{y}>0$. Using
(\ref{eq:V0bound}), the Lyapunov derivative in (\ref{eq:V0dot2})
can be expressed as 
\begin{align}
\dot{V}_{0} & \leq-v\left\Vert z\right\Vert \leq-\frac{v}{\overline{v}}V_{0}.\label{eq:V0dot3}
\end{align}
In (\ref{eq:V0dot3}), $v=min\left(\underline{k_{x}},\underline{y}k_{\theta}\right)\in\mathbb{R},$
where $\underline{k_{x}}\in\mathbb{R}$ denotes the minimum eigenvalue
of the matrix $k_{x}$. The inequalities in (\ref{eq:V0dot3}) can
be used to conclude that $\left\Vert \tilde{\theta}\left(t\right)\right\Vert \to0$
and $\left\Vert \tilde{x}\left(t\right)\right\Vert \to0$ exponentially
fast. Provided the state trajectory $x\left(t\right)$ is bounded%
\footnote{Remark (\ref{rem:chicompact}) in the subsequent analysis shows that
$x\left(t\right)\in\mathcal{L}_{\infty}$.%
}, $\sup_{t}\left\Vert Y\left(t\right)\right\Vert \in\mathcal{L}_{\infty}$.
From (\ref{eq:xtildedot}), $\left\Vert \dot{\tilde{x}}\left(t\right)\right\Vert \leq\left\Vert Y\left(t\right)\right\Vert \left\Vert \tilde{\theta}\left(t\right)\right\Vert +k_{x}\left\Vert \tilde{x}\left(t\right)\right\Vert $,
and hence, $\left\Vert \dot{\tilde{x}}\left(t\right)\right\Vert \to0$.

The concurrent learning-based observer results in exponential regulation
of the parameter and the state derivative estimation errors. In the
following, the parameter and state derivative estimates are used to
approximately solve the HJB equation in (\ref{eq:HJB}) without the
knowledge of the drift dynamics.

\section{Approximate Optimal Control}

Based on the system identifier developed in Section \ref{sec:System-Identifier},
the BE in (\ref{eq:delta1}) can be approximated as
\begin{equation}
\hat{\delta}\left(x,\hat{u},\hat{V},\hat{\theta}\right)=\nabla_{x}\hat{V}\left(Y\hat{\theta}+g\hat{u}\right)+x^{T}Qx+\hat{u}^{T}R\hat{u}.\label{eq:delta2}
\end{equation}
In the following, the approximate BE in (\ref{eq:delta2}) is used
to obtain an approximate solution to the HJB equation in (\ref{eq:HJB}).

\subsection{Value function approximation}

Approximations to the optimal value function $V^{*}$ and the optimal
policy $u^{*}$ are designed based on neural network (NN)-based representations.
The NN-based representation is facilitated by a temporary assumption
that the state trajectory $x\left(t\right)$ evolves on a compact
subspace $\chi\subset\mathbb{R}^{n}$. The compactness assumption
is common in neural network-based adaptive control (cf. \cite{Lewis1993a,Narendra1990}),
and it is shown in the subsequent stability analysis that the states
evolve on a compact set provided the initial condition $x\left(t_{0}\right)$
is bounded (see Remark \ref{rem:chicompact} in the subsequent stability
analysis). The following standard NN assumption describes a NN-based
representation of the optimal value function.
\begin{assumption}
\label{ass:NN}On the compact set $\chi,$ the optimal value function
$V^{*}$ can be represented using a NN as 
\begin{align}
V^{*} & =W^{*T}\sigma+\epsilon,\label{eq:V*NN}
\end{align}
where $W^{*}\in\mathbb{R}^{L}$ is the ideal weight matrix, which
is bounded above by a known positive constant $\bar{W}$ in the sense
that $\left\Vert W^{*}\right\Vert \leq\bar{W}$, $\sigma:\chi\rightarrow\mathbb{R}^{L}=\begin{bmatrix}\sigma_{1} & \cdots & \sigma_{L}\end{bmatrix}^{T}$
is a bounded continuously differentiable nonlinear activation function
such that $\sigma\left(0\right)=0$ and $\sigma'\left(0\right)=0$,
$L\in\mathbb{N}$ is the number of neurons, and $\epsilon:\chi\rightarrow\mathbb{R}$
is the function reconstruction error such that $\sup_{x\in\chi}\left|\epsilon\left(x\right)\right|\leq\bar{\epsilon}$
and\textbf{ }$\sup_{x\in\chi}\left|\epsilon'\left(x\right)\right|\leq\bar{\epsilon}^{\prime},$
where $\bar{\epsilon},\bar{\epsilon}^{\prime}\in\mathbb{R}$ are known
positive constants.
\end{assumption}
Based on (\ref{eq:V*NN}) a NN-based representation of the optimal
controller is derived as
\begin{equation}
u^{*}=-\frac{1}{2}R^{-1}g^{T}\left(\sigma^{\prime T}W^{*}+\epsilon^{\prime T}\right).\label{eq:u*NN}
\end{equation}
The NN-based approximations to the optimal value function in (\ref{eq:V*NN})
and the optimal policy in (\ref{eq:u*NN}) are defined as
\begin{gather}
\hat{V}\triangleq\hat{W}_{c}^{T}\sigma,\quad\hat{u}\triangleq-\frac{1}{2}R^{-1}g^{T}\sigma^{\prime T}\hat{W}_{a},\label{eq:Vu}
\end{gather}
where $\hat{W}_{c}\left(t\right)\in\mathbb{R}^{L}$ and $\hat{W}_{a}\left(t\right)\in\mathbb{R}^{L}$
are estimates of the ideal weights $W^{*}$. The use of two sets of
weights to estimate the same set of ideal weights is motivated by
the stability analysis and fact that it enables a formulation of the
BE that is linear in the value function weight estimates $\hat{W}_{c}$,
enabling a least squares-based adaptive update law.

Based on (\ref{eq:Vu}), the approximate BE in (\ref{eq:delta2})
can be expressed as
\begin{align}
\hat{\delta}\left(x,\hat{W}_{a},\hat{W}_{c},\hat{\theta}\right) & =\omega^{T}\hat{W}_{c}+x^{T}Qx+\hat{u}^{T}R\hat{u},\label{eq:delta3}
\end{align}
where $\omega\left(x,\hat{\theta},\hat{W}_{a}\right)\triangleq\sigma^{\prime}\left(x\right)\left(\hat{f}\left(x,\hat{\theta}\right)+g\left(x\right)\hat{u}\left(x,\hat{W}_{a}\right)\right)$
is the regressor vector.

\subsection{Learning based on desired behavior}

In traditional RL-based algorithms, the value function estimate and
the policy estimate are updated based on observed data. The use of
observed data to learn the value function naturally leads to a sufficient
exploration condition which demands sufficient richness in the observed
data. In stochastic systems, this is achieved using a randomized stationary
policy (cf. \cite{Mehta.Meyn2009,Konda2004,Sutton1998}), whereas
in deterministic systems, a probing noise is added to the derived
control law (cf. \cite{Bhasin.Kamalapurkar.ea2013a,Vamvoudakis2009,Vamvoudakis2010,Dierks2009a,Dierks2010}).
The technique developed in this result is based on the observation
that if an estimate of the system dynamics is available, an approximation
to the BE can be evaluated at any desired point in the state space.
The following condition, similar to the condition in (\ref{eq:RankCond}),
enables the use of approximate BE evaluated at a pre-sampled set of
points $\left\{ x_{i}\in\chi\mid i=1,\cdots,N\right\} $ in the state
space.
\begin{assumption}
\label{ass:ADPLearnCond}There exists a set of points $\left\{ x_{i}\in\chi\mid i=1,\cdots,N\right\} $
such that $\forall t\in\left[0,\infty\right),$ 
\begin{equation}
rank\left(\sum_{i=1}^{N}\frac{\omega_{i}\omega_{i}^{T}}{\rho_{i}}\right)=L.\label{eq:PEFreeCond}
\end{equation}
In (\ref{eq:PEFreeCond}), $\rho_{i}\triangleq1+\nu\omega_{i}^{T}\Gamma\omega_{i}\in\mathbb{R}$
are the normalization terms, where $\nu\in\mathbb{R}$ is a constant
positive normalization gain, $\Gamma:\left(t\right)\in\mathbb{R}^{L\times L}$
is the least-squares gain matrix, and 
\[
\omega_{i}\left(x_{i},\hat{\theta},\hat{W}_{a}\right)\triangleq\sigma^{\prime}\left(x_{i}\right)\left(\hat{f}\left(x_{i},\hat{\theta}\right)+g\left(x_{i}\right)\hat{u}\left(x_{i},\hat{W}_{a}\right)\right).
\]

\end{assumption}
To facilitate the stability analysis, let 
\begin{equation}
\underline{c}=\frac{1}{N}\left(\inf_{t\in[0,\infty)}\left(\lambda_{min}\left\{ \sum_{i=1}^{N}\frac{\omega_{i}\omega_{i}^{T}}{\rho_{i}}\right\} \right)\right),\label{eq:cbar}
\end{equation}
where $\lambda_{min}\left\{ \cdot\right\} $ denotes the minimum eigenvalue.
Provided Assumption \ref{ass:ADPLearnCond} holds, the infimum in
(\ref{eq:cbar}) is positive. The condition in (\ref{eq:RankCond})
is weaker than the PE condition in previous results such as \cite{Bhasin.Kamalapurkar.ea2013a,Vamvoudakis2009,Vamvoudakis2010,Dierks2009a,Dierks2010}
and unlike the PE condition, (\ref{eq:RankCond}) can be verified
online. Since the rank condition in (\ref{eq:PEFreeCond}) depends
on the estimates $\hat{\theta}\left(t\right)$ and $\hat{W}_{a}\left(t\right)$,
it is in general, impossible to guarantee a priori. However, heuristically,
the condition in (\ref{eq:PEFreeCond}) can be met by collecting redundant
data, i.e., by selecting more points than the number of neurons by
choosing $N\gg L$.

The approximate BE can be evaluated at the sampled points $\left\{ x_{i}\mid i=1,\cdots,N\right\} $
as 
\begin{equation}
\hat{\delta}_{i}\left(x_{i},\hat{W}_{a},\hat{W}_{c},\hat{\theta}\right)=\omega_{i}^{T}\hat{W}_{c}+x_{i}^{T}Qx_{i}+\hat{u}_{i}^{T}R\hat{u}_{i},\label{eq:deltai1}
\end{equation}
where $\hat{u}_{i}\left(x_{i},\hat{W}_{a}\right)\triangleq-\frac{1}{2}R^{-1}g\left(x_{i}\right)^{T}\sigma\left(x_{i}\right)^{\prime T}\hat{W}_{a}$.
A concurrent learning-based least-squares update law for the value
function weights is designed based on the subsequent stability analysis
as 
\begin{align}
\dot{\hat{W}}_{c} & =-\eta_{c1}\Gamma\frac{\omega}{\rho}\hat{\delta}-\frac{\eta_{c2}}{N}\Gamma\sum_{i=1}^{N}\frac{\omega_{i}}{\rho_{i}}\hat{\delta}_{i},\nonumber \\
\dot{\Gamma} & =\left(\beta\Gamma-\eta_{c1}\Gamma\frac{\omega\omega^{T}}{\rho}\Gamma\right)\mathbf{1}_{\left\{ \left\Vert \Gamma\right\Vert \leq\overline{\Gamma}\right\} },\:\left\Vert \Gamma\left(t_{0}\right)\right\Vert \leq\overline{\Gamma},\label{eq:criticupdate}
\end{align}
where $\mathbf{1}_{\left\{ \cdot\right\} }$ denotes the indicator
function, $\overline{\Gamma}>0\in\mathbb{R}$ is the saturation constant,
$\beta>0\in\mathbb{R}$ is the forgetting factor, and $\eta_{c1},\eta_{c2}>0\in\mathbb{R}$
are constant adaptation gains. The update law in (\ref{eq:criticupdate})
ensures that the adaptation gain matrix is bounded such that 
\begin{equation}
\underline{\Gamma}\leq\left\Vert \Gamma\left(t\right)\right\Vert \leq\overline{\Gamma},\:\forall t\in[0,\infty)\label{eq:Gammabound}
\end{equation}
The policy weights are then updated to follow the value function weights
as
\begin{align}
\dot{\hat{W}}_{a} & =-\eta_{a1}\left(\hat{W}_{a}-\hat{W}_{c}\right)-\eta_{a2}\hat{W}_{a}\nonumber \\
 & +\left(\frac{\eta_{c1}G_{\sigma}^{T}\hat{W}_{a}\omega^{T}}{4\rho}+\sum_{i=1}^{N}\frac{\eta_{c2}G_{\sigma i}^{T}\hat{W}_{a}\omega_{i}^{T}}{4N\rho_{i}}\right)\hat{W}_{c},\label{eq:actorupdate}
\end{align}
where $\eta_{a1},\eta_{a2}\in\mathbb{R}$ are positive constant adaptation
gains and $G_{\sigma}\triangleq\sigma^{\prime}gR^{-1}g^{T}\sigma^{\prime T}\in\mathbb{R}^{L\times L}$. 

The update law in (\ref{eq:criticupdate}) is fundamentally different
from the concurrent learning adaptive update in results such as \cite{Chowdhary.Johnson2011a,Chowdhary.Yucelen.ea2012},
in the sense that the points $\left\{ x_{i}\in\chi\mid i=1,\cdots,N\right\} $
are selected a priori based on prior information about the desired
behavior of the system. For example, in the present case, since the
objective is to regulate the system states to the origin and the system
is deterministic, it is natural to select a bounded set of points
uniformly distributed around the origin of the state space. This difference
is a result of the fact that the developed RL-based technique uses
the approximate BE as the metric to update the weight estimates. Given
the system dynamics, or an estimate of the system dynamics, the approximate
BE can be evaluated at any desired point in the state space, whereas
in adaptive control, the prediction error is used as a metric which
can only be evaluated at observed data points along the state trajectory.

\section{\label{sec:Stability-analysis}Stability analysis}

To facilitate the subsequent stability analysis, the approximate BE
is expressed in terms of the weight estimation errors $\tilde{W}_{c}\triangleq W^{*}-\hat{W}_{c}$
and $\tilde{W}_{a}\triangleq W^{*}-\hat{W}_{a}$. Subtracting (\ref{eq:HJB})
from (\ref{eq:delta3}), the unmeasurable form of the instantaneous
BE can be expressed as 
\begin{align}
\hat{\delta} & =\nabla_{x}\hat{V}\left(\hat{f}+g\hat{u}\right)+\hat{u}^{T}R\hat{u}-\nabla_{x}V^{*}\left(f+gu^{*}\right)-u^{*T}Ru^{*},\nonumber \\
 & =-\omega^{T}\tilde{W}_{c}-W^{*T}\sigma^{\prime}Y\tilde{\theta}+\frac{1}{4}\tilde{W}_{a}^{T}G_{\sigma}\tilde{W}_{a}+\frac{1}{4}G_{\epsilon}-\epsilon^{\prime}f\nonumber \\
 & +\frac{1}{2}W^{*T}\sigma^{\prime}G\epsilon^{\prime T},\label{eq:delta4}
\end{align}
where $G\triangleq gR^{-1}g^{T}\in\mathbb{R}^{n\times n}$ and $G_{\epsilon}\triangleq\epsilon^{\prime}G\epsilon^{\prime T}\in\mathbb{R}$.
Similarly, the approximate BE evaluated at the sampled states $\left\{ x_{i}\mid i=1,\cdots,N\right\} $
can be expressed as 
\begin{align}
\hat{\delta}_{i} & =-\omega_{i}^{T}\tilde{W}_{c}+\frac{1}{4}\tilde{W}_{a}^{T}G_{\sigma i}\tilde{W}_{a}-W^{*T}\sigma_{i}^{\prime}Y_{i}\tilde{\theta}+\Delta_{i},\label{eq:deltai2}
\end{align}
where $\Delta_{i}\triangleq\frac{1}{2}W^{*T}\sigma_{i}^{\prime}G_{i}\epsilon_{i}^{\prime T}+\frac{1}{4}G_{\epsilon i}-\epsilon_{i}^{\prime}f_{i}\in\mathbb{R}$
is a constant. 

On the compact set $\chi$ the functions $f$ and $Y$ are Lipschitz
continuous, and hence, there exist positive constants $L_{f}$, $L_{Y}\in\mathbb{R}$
such that%
\footnote{The Lipschitz property is exploited here for clarity of exposition.
The bounds in (\ref{eq:Lipschitz}) can be easily generalized to $\left\Vert f\left(x\right)\right\Vert \leq L_{f}\left(\left\Vert x\right\Vert \right)\left\Vert x\right\Vert $,
$\left\Vert Y\left(x\right)\right\Vert \leq L_{Y}\left(\left\Vert x\right\Vert \right)\left\Vert x\right\Vert $,
where $L_{f},\: L_{Y}:\mathbb{R}\to\mathbb{R}$ are positive, nondecreasing
functions.%
} 
\begin{equation}
\left\Vert f\left(x\right)\right\Vert \leq L_{f}\left\Vert x\right\Vert ,\:\left\Vert Y\left(x\right)\right\Vert \leq L_{Y}\left\Vert x\right\Vert ,\forall x\in\chi.\label{eq:Lipschitz}
\end{equation}
Using (\ref{eq:Gammabound}), the normalized regressor $\frac{\omega}{\rho}$
can be bounded as 
\begin{equation}
\overline{\left\Vert \frac{\omega\left(x\right)}{\rho\left(x\right)}\right\Vert }\leq\frac{1}{2\sqrt{\nu\underline{\Gamma}}}.\label{eq:RegBound}
\end{equation}
where the operator $\overline{\left(\cdot\right)}:\left[0,\infty\right)\to\left[0,\infty\right)$
is defined as $\overline{\left(\cdot\right)}\triangleq\sup_{x\in\mathbb{R}^{n}}\left(\cdot\right)$.
The following positive constants are defined to facilitate the subsequent
analysis 
\begin{gather*}
\vartheta_{1}\triangleq\frac{\eta_{c1}L_{f}\overline{\epsilon^{\prime}}}{4\sqrt{\nu\underline{\Gamma}}},\quad\vartheta_{2}\triangleq\sum_{i=1}^{N}\left(\frac{\eta_{c2}\left\Vert \sigma_{i}^{\prime}Y_{i}\right\Vert \overline{W}}{4N\sqrt{\nu\underline{\Gamma}}}\right),\\
\vartheta_{3}\triangleq\frac{L_{Y}\eta_{c1}\overline{W}\overline{\left\Vert \sigma^{\prime}\right\Vert }}{4\sqrt{\nu\underline{\Gamma}}},\quad\vartheta_{4}\triangleq\overline{\left\Vert \frac{1}{4}G_{\epsilon}\right\Vert },\\
\vartheta_{5}\triangleq\overline{\left\Vert \frac{\eta_{c1}\omega\left(2W^{T}\sigma^{\prime}G\epsilon^{\prime T}+G_{\epsilon}\right)}{4\rho}+\sum_{i=1}^{N}\frac{\eta_{c2}\omega_{i}\Delta_{i}}{N\rho_{i}}\right\Vert },\\
\vartheta_{6}\triangleq\overline{\left\Vert \frac{1}{2}W^{T}G_{\sigma}+\frac{1}{2}\epsilon^{\prime}G^{T}\sigma^{\prime T}\right\Vert }+\vartheta_{7}\overline{W}^{2}+\eta_{a2}\overline{W},\\
\vartheta_{7}\triangleq\frac{\eta_{c1}\overline{\left\Vert G_{\sigma}\right\Vert }}{8\sqrt{\nu\underline{\Gamma}}}+\sum_{i=1}^{N}\left(\frac{\eta_{c2}\left\Vert G_{\sigma i}\right\Vert }{8N\sqrt{\nu\underline{\Gamma}}}\right).
\end{gather*}
The main result of this paper can now be stated as follows.
\begin{thm}
\label{thm:mainthm}Provided Assumptions (\ref{ass:fg}) - (\ref{ass:ADPLearnCond})
hold and gains $\underline{q}$, \textup{$\eta_{c2}$}, $\eta{}_{a2}$,
and \textup{$k_{\theta}$} are selected large enough based on the
following sufficient conditions
\begin{gather}
\eta_{a2}>-\frac{\eta_{a1}}{2}+\vartheta_{7}\overline{W}\left(\frac{2\zeta_{2}+1}{2\zeta_{2}}\right),\quad k_{\theta}>\frac{\vartheta_{2}+\zeta_{1}\vartheta_{3}\overline{Z}}{\underline{y}\zeta_{1}},\nonumber \\
\underline{q}>\vartheta_{1},\quad\eta_{c2}>\frac{\zeta_{2}\vartheta_{7}\overline{W}+\eta_{a1}+2\left(\vartheta_{1}+\zeta_{1}\vartheta_{2}+\vartheta_{3}\overline{Z}\right)}{2\underline{c}},\label{eq:conditions1}
\end{gather}
where $\overline{Z}\left(t_{0}\right)\in\mathbb{R}$ is a positive
constant that depends on the initial condition of the system, the
observer in (\ref{eq:observer}) along with the adaptive update law
in (\ref{eq:ThetaUpdate}) and the controller in (\ref{eq:Vu}) along
with the adaptive update laws in (\ref{eq:criticupdate}) and (\ref{eq:actorupdate})
ensure that the state $x\left(t\right)$, the state estimation error
$\tilde{x}\left(t\right)$, the parameter estimation error $\tilde{\theta}\left(t\right)$,
the value function weight estimation error $\tilde{W}_{c}\left(t\right)$
and the policy weight estimation error $\tilde{W}_{a}\left(t\right)$
are UUB, resulting in UUB convergence of the policy $\hat{u}\left(x\left(t\right),\hat{W}_{a}\left(t\right)\right)$
to the optimal policy $u^{*}\left(x\left(t\right)\right)$. \end{thm}
\begin{IEEEproof}
Let $V_{L}:\mathbb{R}^{2n+2L+p}\times\left[0,\infty\right)\to\left[0,\infty\right)$
be a continuously differentiable positive definite candidate Lyapunov
function defined as
\begin{equation}
V_{L}\triangleq V^{*}+\frac{1}{2}\tilde{W}_{c}^{T}\Gamma^{-1}\tilde{W}_{c}+\frac{1}{2}\tilde{W}_{a}^{T}\tilde{W}_{a}+V_{0},\label{eq:VL}
\end{equation}
where $V^{*}$ is the optimal value function, and $V_{0}$ was introduced
in (\ref{eq:V0}). Using the fact that $V^{*}$ is positive definite,
(\ref{eq:V0bound}), (\ref{eq:Gammabound}) and Lemma 4.3 from \cite{Khalil2002}
yield 
\begin{equation}
\underline{v_{l}}\left(\left\Vert Z\right\Vert \right)\leq V_{L}\left(Z,t\right)\leq\overline{v_{l}}\left(\left\Vert Z\right\Vert \right),\label{eq:VLBound}
\end{equation}
for all $t\in\left[0,\infty\right)$ and for all $Z\in\mathbb{R}^{2n+2L+p}$.
In (\ref{eq:VLBound}), $\underline{v_{l}},\overline{v_{l}}:\left[0,\infty\right]\rightarrow\left[0,\infty\right)$
are class $\mathcal{K}$ functions and 
\[
Z\triangleq\left[x^{T},\:\tilde{W}_{c}^{T},\:\tilde{W}_{a}^{T},\:\tilde{x}^{T},\:\tilde{\theta}^{T}\right]^{T}.
\]
The time derivative of (\ref{eq:VL}) along the trajectories of (\ref{eq:dynamics}),
(\ref{eq:xtildedot}), (\ref{eq:ThetaHatDotAlt}), (\ref{eq:criticupdate}),
and (\ref{eq:actorupdate}) is given by
\begin{align*}
\dot{V}_{L} & =\dot{V}^{*}-\tilde{W}_{c}^{T}\Gamma^{-1}\dot{\hat{W}}_{c}-\tilde{W}_{c}^{T}\Gamma^{-1}\dot{\Gamma}\Gamma^{-1}\tilde{W}_{c}-\tilde{W}_{a}^{T}\dot{\hat{W}}_{a}+\dot{V}_{0},\\
 & =\nabla_{x}V^{*}\left(f+g\hat{u}\right)-\tilde{W}_{c}^{T}\left(-\eta_{c1}\frac{\omega}{\rho}\hat{\delta}-\frac{\eta_{c2}}{N}\sum_{i=1}^{N}\frac{\omega_{i}}{\rho_{i}}\hat{\delta}_{i}\right)\\
 & -\frac{1}{2}\tilde{W}_{c}^{T}\Gamma^{-1}\left(\beta\Gamma-\eta_{c1}\left(\Gamma\frac{\omega\omega^{T}}{\rho}\Gamma\right)\right)\Gamma^{-1}\tilde{W}_{c}-\tilde{x}^{T}k_{x}\tilde{x}\\
 & -\tilde{W}_{a}^{T}\left(-\eta_{a1}\left(\hat{W}_{a}-\hat{W}_{c}\right)-\eta_{a2}\hat{W}_{a}\right)\\
 & -\tilde{W}_{a}^{T}\left(\frac{\eta_{c1}G_{\sigma}^{T}\hat{W}_{a}\omega^{T}}{4\rho}+\sum_{i=1}^{N}\frac{\eta_{c2}G_{\sigma i}^{T}\hat{W}_{a}\omega_{i}^{T}}{4N\rho_{i}}\right)\hat{W}_{c}\\
 & -k_{\theta}\tilde{\theta}^{T}\left(\sum_{j=1}^{N}Y_{j}^{T}Y_{j}\right)\tilde{\theta}.
\end{align*}
Substituting for the approximate BEs from (\ref{eq:delta4}) and (\ref{eq:deltai2}),
using the bounds in (\ref{eq:Lipschitz}) and (\ref{eq:RegBound}),
and using Young's inequality, the Lyapunov derivative can be upper-bounded
as
\begin{align*}
\dot{V}_{L} & \leq-\iota_{x}\left\Vert x\right\Vert ^{2}-\iota_{c}\left\Vert \tilde{W}_{c}\right\Vert ^{2}-\iota_{a}\left\Vert \tilde{W}_{a}\right\Vert ^{2}-\underline{k_{x}}\left\Vert \tilde{x}\right\Vert ^{2}\\
 & -\iota_{\theta}\left\Vert \tilde{\theta}\right\Vert ^{2}+\vartheta_{5}\left\Vert \tilde{W}_{c}\right\Vert +\vartheta_{6}\left\Vert \tilde{W}_{a}\right\Vert +\vartheta_{4},
\end{align*}
where $\iota_{x}$, $\iota_{a}$, $\iota_{c}\left(t\right)$, $\iota_{\theta}\left(t\right)\in\mathbb{R}$
are defined as 
\begin{gather*}
\iota_{x}\triangleq\underline{q}-\vartheta_{1},\quad\iota_{a}\triangleq\frac{\eta_{a1}}{2}+\eta_{a2}-\vartheta_{7}\overline{W}\left(\frac{2\zeta_{2}+1}{2\zeta_{2}}\right),\\
\iota_{c}\left(t\right)\triangleq\eta_{c2}\underline{c}-\vartheta_{1}-\zeta_{1}\vartheta_{2}-\frac{\zeta_{2}\vartheta_{7}\overline{W}+\eta_{a1}}{2}-\vartheta_{3}\left\Vert x\left(t\right)\right\Vert ,\\
\iota_{\theta}\left(t\right)\triangleq k_{\theta}\underline{y}-\frac{\vartheta_{2}}{\zeta_{1}}-\vartheta_{3}\left\Vert x\left(t\right)\right\Vert .
\end{gather*}
Provided the conditions in (\ref{eq:conditions1}) are satisfied,
the inequalities 
\begin{gather}
\underline{q}>\vartheta_{1},\quad\eta_{a2}>-\frac{\eta_{a1}}{2}+\vartheta_{7}\overline{W}\left(\frac{2\zeta_{2}+1}{2\zeta_{2}}\right),\nonumber \\
\eta_{c2}>\frac{\zeta_{2}\vartheta_{7}\overline{W}+\eta_{a1}+2\left(\vartheta_{1}+\zeta_{1}\vartheta_{2}+\vartheta_{3}\left\Vert x\left(t\right)\right\Vert \right)}{2\underline{c}},\nonumber \\
k_{\theta}>\frac{\vartheta_{2}}{\underline{y}\zeta_{1}}+\frac{\vartheta_{3}\left\Vert x\left(t\right)\right\Vert }{\underline{y}}\label{eq:conditions2}
\end{gather}
hold for all $t\in\left[0,\infty\right)$ (see Remark \ref{rem:chicompact}),
and hence, the coefficients $\iota_{x}$, $\iota_{a}$, $\iota_{c}\left(t\right)$,
and $\iota_{\theta}\left(t\right)$ are positive for all $t\in\left[0,\infty\right)$.
Completing the squares, the Lyapunov derivative can be expressed as
\begin{align}
\dot{V}_{L} & \leq-\iota_{x}\left\Vert x\right\Vert ^{2}-\frac{\underline{\iota_{c}}}{2}\left\Vert \tilde{W}_{c}\right\Vert ^{2}-\frac{\iota_{a}}{2}\left\Vert \tilde{W}_{a}\right\Vert ^{2}-\underline{k_{x}}\left\Vert \tilde{x}\right\Vert ^{2}\nonumber \\
 & -\underline{\iota_{\theta}}\left\Vert \tilde{\theta}\right\Vert ^{2}+\iota,\nonumber \\
 & \leq-v_{l}\left\Vert Z\right\Vert ,\quad\forall\left\Vert Z\right\Vert \geq\iota>0,\label{eq:VLdot}
\end{align}
where $\iota\triangleq\frac{\vartheta_{5}^{2}}{2\underline{\iota_{c}}}+\frac{\vartheta_{6}^{2}}{2\iota_{a}}+\vartheta_{4}\in\mathbb{R}$
and $\underline{\iota_{c}}$,$\underline{\iota_{\theta}}\in\mathbb{R}$
are the lower bounds on $\iota_{c}\left(t\right)$ and $\iota_{\theta}\left(t\right)$,
respectively. Theorem 4.18 in \cite{Khalil2002} can now be invoked
to conclude that $Z\left(t\right)$ is UUB. \end{IEEEproof}
\begin{rem}
\label{rem:chicompact}If $\left\Vert Z\left(0\right)\right\Vert \geq\iota$
then $\dot{V}_{L}\left(Z\left(0\right)\right)<0$. Thus, $V_{L}\left(Z\left(t\right)\right)$
is decreasing at $t=0$, and hence, $Z\left(t\right)\in\mathcal{L}_{\infty}$
at $t=0^{+}$. Thus all the conditions of Theorem \ref{thm:mainthm}
are satisfied at $t=0^{+}$. As a result, $V_{L}\left(Z\left(t\right)\right)$
is decreasing at $t=0^{+}$. By induction, $\left\Vert Z\left(0\right)\right\Vert \geq\iota\implies V_{L}\left(Z\left(t\right)\right)\leq V_{L}\left(Z\left(0\right)\right),\forall t\in\mathbb{R}^{+}$.
Thus, from (\ref{eq:VLBound}), $\left\Vert Z\left(t\right)\right\Vert \leq\underline{v_{l}}^{-1}\left(\overline{v_{l}}\left(\left\Vert Z\left(0\right)\right\Vert \right)\right),\forall t\in\left[0,\infty\right)$.
If $\left\Vert Z\left(0\right)\right\Vert <\iota$ then (\ref{eq:VLBound})
and (\ref{eq:VLdot}) can be used to determine that $\underline{v_{l}}\left(\left\Vert Z\left(t\right)\right\Vert \right)\leq V_{L}\left(Z\left(t\right)\right)\leq\overline{v_{l}}\left(\left\Vert \iota\right\Vert \right),\forall t\in\left[0,\infty\right)$.
As a result, $\left\Vert Z\left(t\right)\right\Vert \leq\underline{v_{l}}^{-1}\left(\overline{v_{l}}\left(\iota\right)\right),\forall t\in\left[0,\infty\right).$
Let the positive constant $\overline{Z}\in\mathbb{R}$ be defined
as 
\[
\overline{Z}\triangleq\underline{v_{l}}^{-1}\left(\overline{v_{l}}\left(\max\left(\left\Vert Z\left(0\right)\right\Vert ,\iota\right)\right)\right).
\]
This relieves the compactness assumption in the sense that the compact
set $\chi\subset\mathbb{R}^{n}$ that contains the system trajectories
$x\left(t\right),\forall t\in\left[0,\infty\right)$ is given by $\chi\triangleq\left\{ x\in\mathbb{R}^{n}\mid\left\Vert x\right\Vert \leq\overline{Z}\right\} .$
Furthermore, $\left\Vert Z\left(t\right)\right\Vert \leq\overline{Z},\forall t\in\left[0,\infty\right)$
implies that the gain conditions in (\ref{eq:conditions1}) are sufficient
for the inequalities in (\ref{eq:conditions2}) to hold for all $t\in\left[0,\infty\right).$
\end{rem}

\section{Conclusion}

A RL-based online approximate optimal controller is developed that
does not require PE for convergence. The PE condition is replaced
by a weaker rank condition that can be verified online from recorded
data. An approximation to the BE computed at pre-sampled desired values
of the system state using an estimate of the system dynamics is used
to improve the value function approximation, and UUB convergence of
the system states to the origin, and UUB convergence of the policy
to the optimal policy are established using a Lyapunov-based analysis.

\bibliographystyle{IEEEtran}
\bibliography{ncr,master,encr}

\end{document}